\documentstyle[12pt]{article}
\begin{document}
\begin{center}
{\bf{\large{Local Cosmic String In Generalised Scalar Tensor Theory}}}\\
\vskip 20pt

A.A.Sen\footnote{E-mail:anjan@juphys.ernet.in},N.Banerjee\footnote{narayan@
juphys.ernet.in}\\

Relativity and Cosmology Research Centre\\
Department of Physics, Jadavpur University\\
Calcutta 700032, India.\\
\end{center}

\centerline{\underline{Abstract}}

A recent investigation shows that a local gauge string with a phenomenological
energy momentum tensor, as prescribed by Vilenkin, is inconsistent in Brans-Dicke
theory. In this work it has been shown that such a string is consistent in a more 
general scalar tensor theory where $\omega$ is function of the scalar field.A
set of solutions of full nonlinear Einstein's equations for interior 
region of such a string are presented.
\newpage

\par
In a very recent communication \cite{R1}, the present authors have shown that an infinitely
long straight static local gauge string, given by the energy momentum tensor 
components $T^{t}_{t}=T^{z}_{z}\neq0$ and all other $T^{\mu}_{\nu}=0$ \cite{R2},
 is inconsistent in Brans-Dicke(BD) theory of gravity. Because of the relevence of the BD 
 type scalar 
 field in the context of a cosmic string (see Gundlach and Ortiz \cite{R3} or Romero
 and Barros \cite{R4} for detailed discussions), it becomes necessary to investigate
 whether a local gauge string can give rise to consistent solutions of the gravitational
 field equations for a more general scalar tensor theory. In this  paper, we 
 show that such a string is indeed consistent with Nordtvedt's generalized
 scalar tensor theory \cite{R5}.
 \par
 The gravitational field equations in this theory are given by
 \begin{equation}
 G_{\mu\nu}= {T_{\mu\nu}\over{\phi}}+{\omega(\phi)\over{\phi^{2}}}(\phi_{,\mu}\phi_{,\nu}-
 {1\over{2}}g_{\mu\nu}\phi^{,\alpha}\phi_{,\alpha})+{1\over{\phi}}(\phi_{,\mu;\nu}-
 g_{\mu\nu}\Box\phi),
 \end{equation}
 where the dimensionless parameter $\omega$ is now a function of the scalar field 
 $\phi$. The wave equation for the scalar field $\phi$ is 
 \begin{equation}
 \Box\phi={1\over{(2\omega+3)}}[T-\phi^{,\alpha}\phi_{,\alpha}{d\omega\over{d\phi}}].
 \end{equation}
 In these equations, $T_{\mu\nu}$ represents the energy momentum tensor components
 for all the fields except the scalar field $\phi$ and $T$ is the trace of $T_{\mu\nu}$.
 Brans Dicke theory is evidently a special case of this theory when $\omega $
 is constant. The conservation of matter is represented by the equation
 \begin{equation}
 T^{\mu\nu}_{;\nu}=0.
 \end{equation}
 It should be noted, however, the equation (2) and (3) are not independent, as in
 view of the equation(1) and the Bianchi identity, one yields the other.

 The general static cylindrically symmetric metric,
 \begin{equation}
 ds^{2}=e^{2(K-U)}(-dt^{2}+dr^{2})+e^{2U}dz^{2}+e^{-2U}W^{2}d\theta^{2},
 \end{equation}
 is taken to describe the spacetime given by an infinitely long static local string
 with the axis of symmetry being the z-axis. $K,U,W$ are all functions of the
 radial coordinate $r$ alone. The local gauge string is characterised by an
 energy density and a stress along the symmetry axis given by
 \begin{equation}
 T^{t}_{t}=T^{z}_{z}=-\sigma(r),
 \end{equation}
 and all other components are zero \cite{R2}.The field equations
 can be written as 
 \begin{eqnarray}
 -{W^{\prime\prime}\over{W}}+{K^{\prime}W^{\prime}\over{W}}-U^{\prime2}
 &=&{\sigma e^{2(K-U)}\over{\phi}}
 +{\omega\over{2}}{\phi^{\prime2}\over{\phi^{2}}}-(K^{\prime}
 -U^{\prime}){\phi^{\prime}\over{\phi}}\nonumber\\
 &+&({\phi^{\prime\prime}\over{\phi}}+{W^{\prime}\phi^{\prime}\over{W\phi}}),\\
{K^{\prime}W^{\prime}\over{W}}-U^{\prime2}
&=&{\omega\over{2}}{\phi^{\prime2}\over{\phi^{2}}}
-(K^{\prime}-U^{\prime}){\phi^{\prime}\over{\phi}}
-{W^{\prime}\phi^{\prime}\over{W\phi}},\\
K^{\prime\prime}+U^{\prime2}&=&-{\omega\over{2}}{\phi^{\prime2}\over{\phi^{2}}}
-{\phi^{\prime\prime}\over{\phi}}-{U^{\prime}\phi^{\prime}\over{\phi}},\\
 -{W^{\prime\prime}\over{W}}-U^{\prime2}+2U^{\prime\prime}
 +2{U^{\prime}W^{\prime}\over{W}}
 -K^{\prime\prime}&=&{\sigma e^{2(K-U)}\over{\phi}}
+{\omega\over{2}}{\phi^{\prime2}\over{\phi^{2}}}
-{U^{\prime}\phi^{\prime}\over{\phi}}\nonumber\\
&+&({\phi^{\prime\prime}\over{\phi}}+{W^{\prime}\phi^{\prime}\over{W\phi}}),
 \end{eqnarray}
 where a prime represents differentiation with respect to $r$.

 The wave equation for the scalar field now looks like 
 \begin{equation}
 \phi^{\prime\prime}+{\phi^{\prime}W^{\prime}\over{W}}
 =-{2\sigma e^{2(K-U)}\over{(2\omega+3)}}
 -{\phi^{\prime2}\over{(2\omega+3)}}{d\omega\over{d\phi}}.
 \end{equation}
 From the conservation equation(3) one can write
 \begin{equation}
 K^{\prime}\sigma=0.
 \end{equation}
 For a nontrivial existence of the cosmic string, $\sigma \neq 0$, and hence 
 \begin{equation}
 K^{\prime}=0,
 \end{equation}
 for the interior of the string. So $K$ is a constant and in what follows we
 shall  take $e^{2K}=1$ which only leads to rescaling of the coordinates and no loss
 of generality. With $K^{\prime}=0$, equation(7) and (8) now combine to yield the 
 equation
 \begin{equation}
 {\phi^{\prime\prime}\over{\phi}}+{W^{\prime}\phi^{\prime}\over{W\phi}}=0.
 \end{equation}

 In case of BD theory, ${d\omega\over{d\phi}}=0$ and thus equation (13) and    (10)
 together yield $\sigma=0$ indicating the nonexistence of the string \cite{R1}.  
 For a varying $\omega$ theory, however, $\sigma$ does not have to vanish as 
 evident from equation(10).
 For $\phi^{\prime}\neq0$, equation(13) readily integrates to yield
 \begin{equation}
 \phi^{\prime}={a\over{W}},
 \end{equation}
 $a$ being a constant of integration and should be nonzero to have a non
 trivial scalar field.
 In view of equation (12) and (13), the field 
 equations (6) and (9) combine to form
 \begin{equation}
 U^{\prime\prime}+{U^{\prime}W^{\prime}\over{W}}+{U^{\prime}\phi^{\prime}\over{\phi}}=0.
 \end{equation}
 This equation has a first integral,
 \begin{equation}
 U^{\prime}W\phi=a_{1},
 \end{equation}
 when $a_{1}$ is an arbitary constant. This equation along with equation (14) 
 yields,
 $$
 U^{\prime} = b{\phi^{\prime}\over{\phi}},
 \eqno{(17)}$$
 where $b=a_{1}/a$.
 
In what follows, we shall try to find out exact solutions for the interior
spacetime metric (i.e. for $\sigma\neq 0$).

\underline{\bf{Interior solution}}

 As a consequence of equation (12) and (13), equation (7) and (8) become 
 identical and we are left with four independent equations,
 $$
 \sigma = - {1\over{2}}{(\phi^{\prime}\phi^{b})}^{2}{d\omega\over{d\phi}}
 \eqno{(18a)}$$
 $$
 -{\phi^{\prime\prime}\over{\phi}}=(D+{\omega\over{2}}){\phi^{\prime2}\over{\phi^{2}}},
 \eqno{(18b)}$$
 $$
 \phi^{\prime}={a\over{W}}
 \eqno{(18c)}$$
 $$
 U^{\prime} = b{\phi^{\prime}\over{\phi}}
 \eqno{(17)}$$
 where $D=b^{2} +b$, a constant.\\
 Equation (18a) follows from (10) and (13) whereas equation 
 (18b) follows from (8).
 Now one has four equations and five unknowns to be solved 
 from them. But in the 
 generalised scalar tensor theories $\omega$ is a function of $\phi$ and 
 thus for
 a particular choice of $\omega=\omega(\phi)$, 
 the system of equations can be solved.

 A number of different choices of $\omega$ as function of $\phi$ are already 
 available in the literature, depending on particular physical interests. 
 Barkar's choice of $\omega$ \cite{R6}, given by 
 $$
 \omega = {{4-3\phi}\over{2(\phi-1)}}
 \eqno{(19)}$$
 will be used for further analysis in the present work. 
 The physical motivation
 for the choice of Barkar is that $G$, the Newtonian constant of gravitation,
 remains a constant in this case in spite of the nonminimal coupling between
 the scalar field and geometry.

 From equation (19) on obtains
 $$
 {d\omega\over{d\phi}} = -{1\over{2}}{1\over{(\phi-1)^2}}
 \eqno{(20)}$$
 and equation (18a) yields
 $$
 \sigma = {1\over{4}}[{{\phi^{\prime}\phi^{b}}\over{(\phi-1)}}]^{2}
 \eqno{(21)}$$
 which evidently ensures that $\sigma$ is positive. It deserves mention that 
 for some other choices of $\omega$, already available in the literature,
 $\sigma$ turns out to be negative. 
 One such choice is $\omega = {{3\phi}\over{2(1-\phi)}}$ which is called
 the model with curvature coupling[see Van den Bergh \cite{R7} and references
 therein]. Obviously these theories do not incorporate a cosmic string of 
 this type.

 In what follows, we will try to solve the system of equations in Barkar's
 theory for two choices of the constant, b and D; namely $1) b=0, D=0$ and
 $2)D=1$.

{\underline{\bf{Case 1$> D=0, b=0$}}}

In this case, from equation (17), one can find $U^{\prime}=0$, i.e. 
$U$ is a constant, and one can choose $U=0 (i.e. e^{2U}=1)$ without
any loss of generality, by a simple rescaling of the coordinates. Physically
this choice allows a Lorentz boost along the symmetry axis of the string.

Equation (21) yields
$$
\sigma = {1\over{4}}{{\phi^{\prime 2}}\over{(\phi-1)^2}}  
\eqno{(22)}$$

With the help of equation (19), equation (18b) can be written as
$$
 {{\phi^{\prime\prime}}\over{\phi^{\prime}}}=[{1\over{\phi}}
 -{1\over{4(\phi-1)}}]{\phi^{\prime}}
 $$
which readily yields a first integral
 $$
 ln {\phi^{\prime}}= ln{\phi_{0}}[\phi/{(\phi-1)^{1/4}}]
 \eqno{(23)}, $$
where $\phi_{0}$ is a constant of integration. A series solution of this 
 equation is possible, expressing r as a power series of $\phi$, 
 which, however,is not invertible to express $\phi=\phi(r)$. 
 But as $\sigma, \omega, W$ are known functions of $\phi$ and its derivatives, 
 the complete solution can be obtained in principle.

{\underline{\bf{Case 2$> D=1$}}}

For this case equation (18b) together with equation (19) yields a solution 
for $\phi$ in the closed form:
$$
\phi = 1+(mr+n)^{4/5}
\eqno{(24)}$$
where $m$ and $n$ are arbitary integration constant. Equations (18c) and (17)
yield the solutions for $W$ and $U$ respectively as
$$
W^{2} = W_{0}^{2}(mr+n)^{2/5}
\eqno{(25)}$$
and
$$
e^{2U} = \phi^{2b} = [1+(mr+n)^{4/5}]^{2b}
\eqno{(26)}$$
Here $W_{0} = 5a/4m$.\\
The complete solution for the metric for the interior of the string is then
$$
ds^{2}=[1+(mr+n)^{4/5}]^{-2b}[-dt^{2}+dr^{2}+W_{0}^{2}(mr+n)^{2/5}d\theta^{2}]\\
\hspace{4mm}+[1+(mr+n)^{4/5}]^{2b}dz^{2}
\eqno{(27)}$$
and the string energy density $\sigma(r)$ is given by
$$
\sigma={{4m^{2}[1+(mr+n)^{4/5}]^{2b}}\over{25(mr+n)^{2}}}
\eqno{(28)}
$$

As a conclusion, one can say that although the solutions obtained in the present work
are by no means the general ones, but it explicitly exhibits a consistent set of 
interior solutions of the nonlinear Einstein's equations for a local gauge string
in a varying $\omega$ theory.So although a local cosmic string is inconsistent 
in Brans-Dicke theory,
it is quite consistent in a more general scalar tensor theory of gravity.

It deserves mention that in a recent work by Guimar\~{a}es \cite{R8}, the solutions
of Einstein's equations for a gauge string have been presented in a weak field 
approximation
of the field equations in similar $\omega$-varying scalar tensor theory. Guimar\~{a}es'
work does not consider the phenomenological expression for $T^{\mu}_{\nu}$, i.e.
$T^{0}_{0}=T^{z}_{z}\neq0$ and all other $T^{\mu}_{\nu}=0$ as prescribed by Vilenkin,
but rather has all the diagonal components of $T^{\mu}_{\nu}$ to be nonvanishing.
This type of a cosmic string is consistent in BD theory as well, even in the full 
nonlinear version of the theory, as shown by the present authors \cite{R1}.
\vskip 15pt
{\large{\bf{ACKNOWLEDGEMENT}}}

The authors are grateful to Prof.A.Banerjee for his generous help and suggestions.
One of the authers (AAS) is grateful to the University Grants Commission, India,
for the financial support.

 \end{document}